\documentclass[conference,letterpaper]{IEEEtran}
\usepackage[top=0.75in, bottom=1.05in, left=0.625in, right=0.625in]{geometry}
\usepackage[bookmarks=false, pdfborder={0 0 0}, colorlinks=false]{hyperref}
\usepackage{cite}
\usepackage{amsmath,amssymb,amsfonts}
\usepackage{algorithmic}
\usepackage{listings}
\usepackage{graphicx}
\usepackage{textcomp}
\usepackage{xcolor}
\usepackage{soul}
\def\BibTeX{{\rm B\kern-.05em{\sc i\kern-.025em b}\kern-.08em
    T\kern-.1667em\lower.7ex\hbox{E}\kern-.125emX}}

\begin{document}
\pagestyle{empty}

\title{Remote Assistance or Remote Driving: \\ The Impact of Operational Design Domains on ADS-Supporting Systems Selection}

\author{\IEEEauthorblockN{
Ole Hans}
\IEEEauthorblockA{\textit{Department of Operational Safety} \\
\textit{Vay Technology GmbH}\\
Berlin, Germany \\
ole.hans@vay.io}
\and
\IEEEauthorblockN{Benedikt Walter}
\IEEEauthorblockA{\textit{Institute of Aircraft Design} \\
\textit{University of Stuttgart}\\
Stuttgart, Germany \\
walter@ifb.uni-stuttgart.de}
}

\maketitle

\begin{abstract}
High level Automated Driving \mbox{Systems (ADS)} can handle many situations, but they still encounter situations where human intervention is required. In systems where a physical driver is present in the vehicle, typically SAE Level 3 systems, this intervention is relatively straightforward and is handled by the in-vehicle driver. However, the complexity increases for Level 4 systems, where, in most cases, no physical driver remains in the vehicle. The two common industry solutions for this challenge are the integration of a remote support system, such as a Remote Driving \mbox{System (RDS)} or Remote Assistance \mbox{System (RAS)}. While it is clear that ADS will require one of these systems, it is less clear how the suitability of either system for a particular ADS application should be evaluated. Currently, the selection process often focuses on system architecture as well as its design and integration challenges. Furthermore, since many ADS developers choose to develop remote system solutions in-house, it is advantageous to select the simpler approach to streamline development and integration efforts. While these decision points are certainly relevant, this approach overlooks the most critical factors: the use cases and the complementarity of the ADS and the remote support system within the context of the Operational Design Design \mbox{Domain (ODD)}. This paper proposes a structured approach for selecting between RDS and RAS as an ADS support system, based on the defined ODD and use case analysis. To achieve this, the paper applies the PEGASUS framework to systematically describe and analyze the ODD. A structured framework is introduced to evaluate and select the most suitable remote support system for an ADS based on clearly defined criteria.
\end{abstract}


\section{Introduction}
\label{sec:intr}
Automated Driving \mbox{Systems (ADS)} are widely regarded as a key technology for transforming the transportation sector, as they have the potential to enhance efficiency, safety, and driving comfort while reducing both the economic and environmental costs of mobility \cite{schoitsch2016autonomous}. At the same time, ADS face significant challenges, particularly in urban environments where continuous interaction with other road users is required. A central issue in this context is the occurrence of so-called “stranded vehicle” situations, in which ADS encounter traffic scenarios that exceed their current technical capabilities and therefore necessitate human intervention. This challenge is further exacerbated by the system's need to interpret and respond appropriately to the often unpredictable behavior of other traffic participants. As such, these limitations represent a challenge to the full automation of driving tasks and to the broader societal acceptance of such \mbox{systems \cite{mutzenich2021updating}}.

To resolve the potential situation whenever the ADS cannot perform the Dynamic Driving \mbox{Tasks (DDT)} anymore, these systems are equipped with a Minimum Risk \mbox{Maneuver (MRM)} capability that safely stops the vehicle to achieve a Minimum Risk \mbox{Condition (MRC)}. From that MRC, the ADS often is able to resume the DDT. However, in some situations that is not the case, which results in the fact that ADS still depends on supporting systems to handle specific scenarios beyond the capabilities of the ADS itself. Common solutions include the integration of Remote Driving \mbox{System (RDS)} and Remote Assistance \mbox{Systems (RAS)}, which, despite their similarities, are often chosen based on system architecture or short-term availability rather than a thorough evaluation of their suitability for particular ADS use cases. This approach frequently leads to suboptimal product-level outcomes, with certain essential use cases remaining unaddressed.

Current robotaxi projects like Zoox or Waymo have chosen remote assistance solutions. It appears that this choice is made for two major reasons: 

\begin{itemize}
    \item Firstly, RAS integrates smoothly into an existing ADS. The remote operator input merely replaces the perception input. The remaining ADS pipeline remains untouched, and the output is still generated by the ADS rather than the remote operator.
    \item Second, most robotaxi developers have built their own remote systems. Therefore they naturally select the system that is less complex and has lower requirements for hardware, software, and performance. 
\end{itemize}

In contrast, companies like Vay Technology \cite{vay_teledriving} and Roboauto \cite{roboauto_teleoperation_2025} provide remote driving solutions that can be integrated in cars or trucks and allow to fully perform remote operation or support of an ADS system in a hybrid setup. The technology is more demanding than an RAS solution, therefore it is not by accident that there are companies solely focused on developing such systems. The main driver for selection between RAS and RDS therefore appears to be the integration effort and the overall complexity. This does not consider the actual reason why remote systems where needed in the first place. Supporting an ADS in a situation where it fails to perform the DDT due to a complex situation occurring in some scenario within the Operational Design Domain (ODD). Therefore, the authors argue that the primary decision point for the selection between RDS and RAS should be the ODD-related use case demands of an ADS system. While other factors might play a role, the main purpose of a remote technology is to enable an ADS to continue its DDT and therefore it must be evaluated which use cases for a given ODD cannot be performed by the ADS alone. This allows for an informed decision between RDS and RAS based on the technology capabilities and the resulting support for the different use cases. 

This paper presents a structured methodology for evaluating RAS and RDS within the context of the ODD. Based on the ODD definition part of the PEGASUS \mbox{method \cite{wachenfeld2016safety}}, the ODD is systematically analyzed to identify relevant use cases for these fallback systems, and the suitability of each system type is assessed accordingly. The results offer a comparative analysis of RAS and RDS as backup solutions for ADS, providing a foundation for informed decision-making and improved integration into ADS architectures. In the following paper, \mbox{Section \ref{relatedwork}} provides an overview of the current state of research. \mbox{Section \ref{method}} outlines the applied methodology. The core of the paper is presented in \mbox{Section \ref{sec:evaluation}}, where RAS and RDS are evaluated with respect to ODD-specific use cases. This comparison highlights the capabilities and limitations of each system and supports the selection of an appropriate fallback strategy in different ADS scenarios. The findings are discussed in detail in the result \mbox{Section \ref{sec:verANDval}} and concluding a brief outlook on future work.

\section{Literature Research}
\label{relatedwork}
A widespread concept for supporting unmanned vehicles is the concept of remote operation, such as remotely operated vehicles for Mars exploration or in the defense \mbox{sector \cite{diermeyer2011mensch}}. In the civilian sector, unmanned remote vehicles are often used for surveillance \cite{fong2001vehicle} or as fallback solutions for automated driving \mbox{vehicles \cite{majstorovic2022survey}}. With regard to solving the challenges mentioned in \mbox{Section \ref{sec:intr}}, the concept of remote operation is a solution to overcome the situations in which the ADS cannot handle a particular situation by calling for remote support. Remote operation can also intervene when automated driving is not expected to be available on certain sections of the route or at certain times of day. Once such situation occurs a predetermined takeover by remote operation takes place. In the automotive industry, remote operation is subdivided into the concepts of remote driving or remote control, remote assistance, and remote monitoring, as further elaborated in the following Sections.

\begin{figure}[ht]
\centering
  \includegraphics[width=0.48\textwidth]{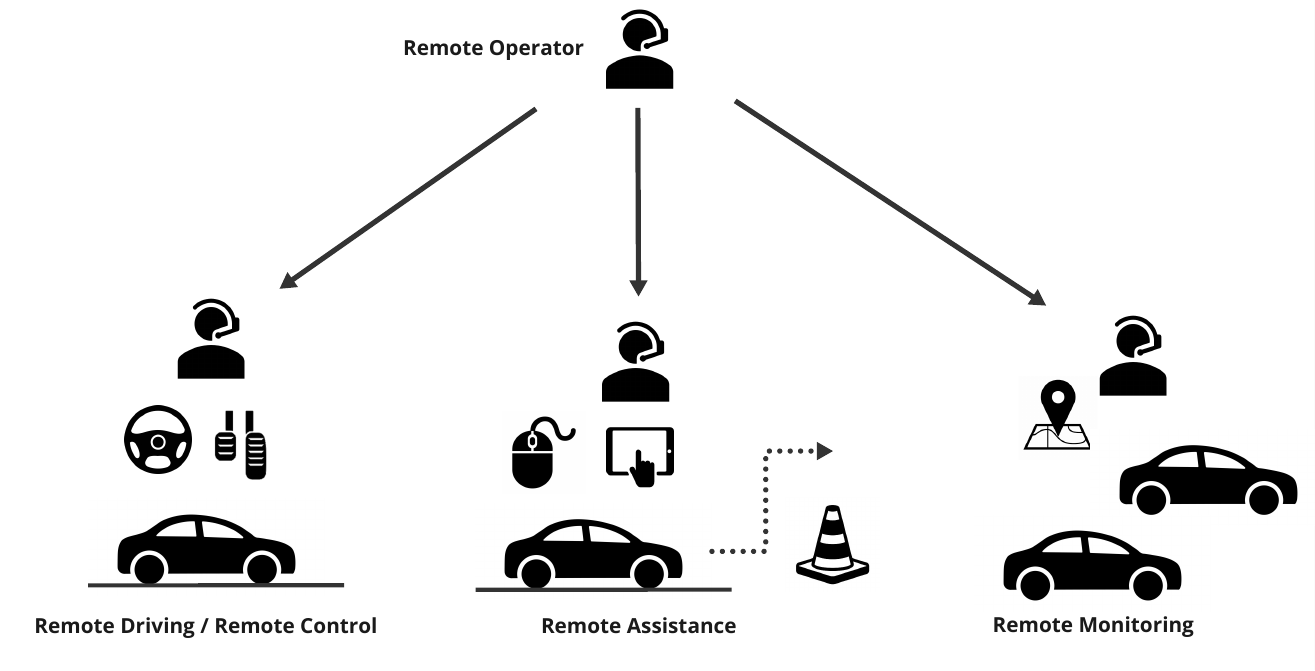}
    \caption{Simplified visualization of Remote Operation Concepts \cite{UNECE, mutzenich2021updating}}
    \label{Fig:.Remote_Operation_concepts}
\end{figure}

\subsection{Remote Monitoring}
\label{sec:remote_monitoring}
In the Remote Monitoring concept, the Remote Operator acts as an observer to monitor the ADS driving behavior remotely in real-time. Accordingly, this is an operating mode in which the Remote Operator only has an indirect influence on the ADS. The Remote Operator is limited to activities like real-time systems diagnostics monitoring and generation of high-level commands to continuously monitor and diagnose the vehicle \mbox{system \cite{fong2001vehicle}}. In Remote Monitoring the Remote Operator does not have the capabilities to take over any of the vehicle's \mbox{DDT \cite{BSIFlex1886}}. Application-related Remote Monitoring systems often transmit non-standardized information to their counterparts, such as information on the condition of the vehicle \mbox{doors \cite{amador2022survey}}. 

\subsection{Remote Assistance}
\label{sec:remote_assistance}
 The concept of Remote Assistance is an \textbf{indirect control} method where the Remote Assistant only intervenes in exceptional cases. This can be differentiated into situations where the ADS reaches its functional \mbox{limits \cite{kettwich2021teleoperation}}, in the event of a vehicle breakdown, or when deciding how to perform certain safety-related tasks. An example is a situation where the ADS encounters unexpected and untrustworthy driving scenarios, such as roadworks or occurring \mbox{obstacles \cite{lu2022teleoperation}}, and remote assistance supports by determining the safest route. According to \mbox{Brecht et al. \cite{brecht2024evaluation}}, a distinction can be made between the different concepts of remote assistance, which are Perception Modification, Collaborative Path Planning, Waypoint Guidance, and Trajectory Guidance. For instance, the german law on autonomous driving introduced the Technical Supervisor as a new legal actor. In accordance with \S 14 Autonome-Fahrzeuge-Genehmigungs-und-Betriebs-\mbox{Verordnung (AFGBV) \cite{AFGBV2022}}, this person is responsible for ensuring compliance with traffic law requirements without permanently monitor driving operations.

\subsection{Remote Driving}
\label{sec:remote_driving}
Remote Driving, also known as remote control, involves direct and full control of the vehicle and refers to real-time control over an extended period of \mbox{time \cite{bogdoll2022taxonomy}}.  The Remote Operator, named as Remote Driver in the context of Remote Driving, performs the entire DDT and \textbf{directly controls} a vehicle through complex traffic maneuvers, making decisions related to steering, acceleration, braking, and route selection without physically being inside the \mbox{vehicle \cite{kettwich2021teleoperation}}. The Remote Driver is a driver who is not seated in a position to manually exercise in-vehicle control commands and transmission gear selection input devices (if any), but is able to operate the \mbox{vehicle \cite{SAEJ3016:2021}} from afar. The remote control technology can be applied in various applications, such as remote-controlled vehicles in urban traffic environments or remote-controlled trucks on private harbor \mbox{sites \cite{amador2022survey}}. The core of remote driving lies in a camera-based in-vehicle system, a stable wireless connection, and a display and control unit at a Remote Control Station. This setup allows the Remote Driver to perceive the vehicle information, traffic environment, and thus part of the ODD through cameras with time-delayed data transmission in the Remote Control \mbox{Station \cite{neumeier2019teleoperation}}. The Remote Driver is able to process these inputs and provide a control output to execute the DDT for the paired vehicle. 

Compared to the direct control approach, the \textbf{shared control} can be regarded as an extended version of direct control with additional safety measures according to \mbox{Schitz et al. \cite{schitz2021interactive}}. In the shared control concept, the Remote Driver's control inputs are evaluated based on various safety objectives, such as collision avoidance. If the Remote Driver's control inputs do not meet these objectives, the vehicle control system, for example an Automated Emergency Braking System overrides the Remote Driver's control inputs and adapts them to achieve the \mbox{objectives \cite{schimpe2022open}}. An evaluation of the shared control concept that assists the Remote Driver can be found in the work of Brecht and \mbox{Diermeyer \cite{brecht2024risk}}.

\subsection{Operational Design Domain}
The ODD is defined as a subset of the operational domain, which encompasses the totality of a geographic region and its environmental conditions over a specified time \mbox{interval \cite{rohne2022implementing}}. As a core concept in the development of ADS, RAS, and RDS, the ODD plays a pivotal role in safety assessment and system validation by ensuring operation within a predefined and predictable range of \mbox{conditions \cite{sun2021acclimatizing}}. If an ADS or RDS can no longer fulfill its ODD-based operational requirements, due to system limitations or failures, a fallback mechanism of the DDT must be triggered. This fallback can either involve a human safety driver or lead to the initiation of an MRM to transition the system into an MRC, often through controlled \mbox{deceleration \cite{SAEJ3016:2021}}.

An ODD constrains the system’s capabilities based on static and dynamic environmental and infrastructural factors, including road type, traffic density, weather, lighting conditions, and other \mbox{parameters \cite{czarnecki2018towards, koopman2019many}}. Given the complexity and high dimensionality of these descriptors, simplified ODD models and taxonomies have been \mbox{developed \cite{PAS, FDIS}}. The PEGASUS \mbox{method \cite{wachenfeld2016safety}}, for instance, introduced a six-layer model for structured scenario-based validation in addition to other best practice approaches for describing an ODD, such as the Automated Vehicle Safety \mbox{Consortium (AVSC) \cite{AVSC}}. This model enables test case derivation by systematically addressing the physical environment, infrastructure, time-dependent changes, dynamic objects, environmental conditions, and digital \mbox{infrastructure \cite{wachenfeld2016safety}}. A systematic approach for defining and analyzing an ODD for RDSs in an urban environment has been outlined by \mbox{Hans et al. \cite{hans2023operational}}.

Building upon these foundations, Hans and \mbox{Walter \cite{hans2024backedautonomy}} analyzed the ODD definitions for ADS and RDS using the PEGASUS layer model and showed systematic differences in the limitations of both systems. While ADS is especially limited by hard-to-predict events such as temporary changes (layer 3) and dynamic objects (layer 4), RDS overcomes these challenges better through real-time human decision making. Conversely, RDS is highly dependent on stable network connectivity (layer 6), which can lead to fragmentation of the ODD, also called "ODD-Islands" \cite{hans2024backedautonomy}, which is a problem that does not affect ADS due to local autonomy. Based on these complementary limitations, the authors propose a combined ODD design in which ADS and RDS act together in hybrid systems. In this way, ADS can take over in areas with weak infrastructure, while RDS secures complex or ambiguous traffic situations. This combination enables effective ODD expansion and increases the operational availability of automated mobility systems.

\section{Methodology}
\label{method}
\begin{figure}[ht]
\centering
  \includegraphics[width=0.48\textwidth]{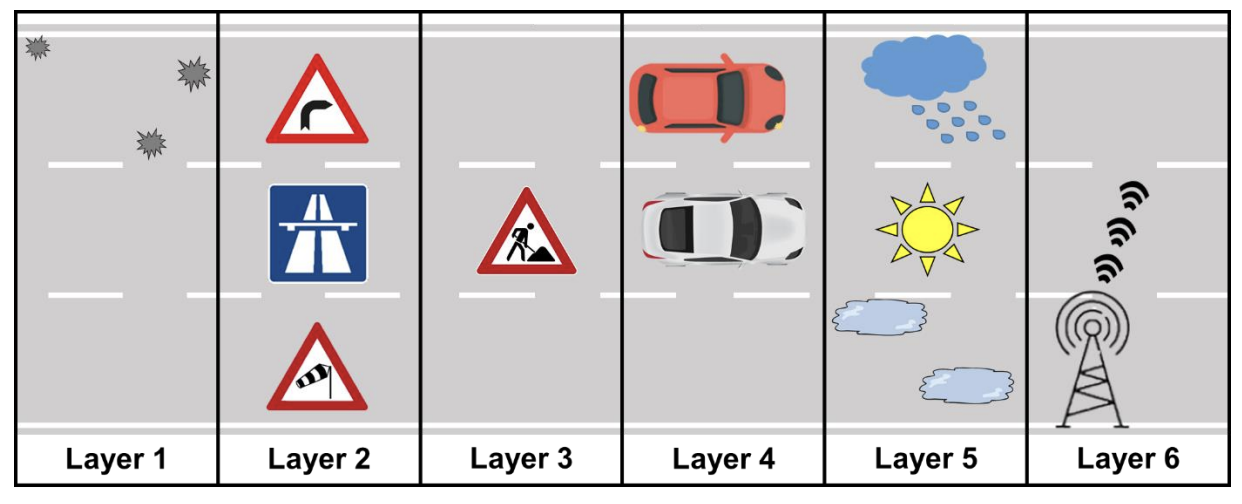}
    \caption{Simplified visualization of the six PEGASUS layers.}
    \label{PegasusLayer}
\end{figure}

The PEGASUS \mbox{method \cite{wachenfeld2016safety}} is used for the structured derivation and evaluation of the hypotheses set out in this work. This method offers a systematic approach to defining and analyzing the ODD and is based on a six-layer model, as shown in \mbox{Fig. \ref{PegasusLayer}}. 

The structured approach based on the PEGASUS layers enables a targeted derivation of hypotheses that address central aspects of ODD design and their effects on the performance and safety of remote support systems. The following hypotheses are discussed in the evaluation of this work.

\begin{itemize}
    \item \textbf{Hypothesis 1:} RDSs are better suited than RASs to cope with dynamic and temporary changes in the traffic situation, such as roadworks in layer 3.
    \item \textbf{Hypothesis 2:} In layer 5, which includes both dynamic and strategic aspects (e.g. weather, visibility, strategic route planning), the suitability of RDSs or RASs depends on the specific use case where RDS is more suitable for dynamic interventions while driving, whereas RAS is more suitable for strategic decisions such as route adjustment.
    \item \textbf{Hypothesis 3:} RASs are more advantageous in layer 6 than RDSs, as they do not require direct control and real-time interaction like RDSs. Consequently, their demands on bandwidth and latency are lower, making them more robust in environments with limited or fluctuating network availability.
    \item \textbf{Hypothesis 4:} RDSs are better suited for longer distances because they allow direct remote control, whereas RASs do not offer this capability and are therefore limited to situations within shorter ranges.
\end{itemize}

\section{Evaluation}
\label{sec:evaluation}
The PEGASUS ODD definition framework is applied by going through layer by layer with the intend to point out ADS limits. Further, RDS and RAS are compared for their suitability to support ADS in the use cases where ADS cannot resolve the issue as a stand-alone technology today. This approach will ultimately provide answers to the hypothesis raised in \mbox{Section \ref{method}}. In order to discuss this with a suitable use case, the authors assume a SAE \mbox{Level }4 ADS in a semi-truck and a hub-to-hub delivery use case with primarily driving on highways and limited exposure to surface streets between the hub and the highway, such as Aurora, Kodiak Robotics, or Torc Robotics. 

\begin{figure}[ht]
\centering
  \includegraphics[width=0.35\textwidth]{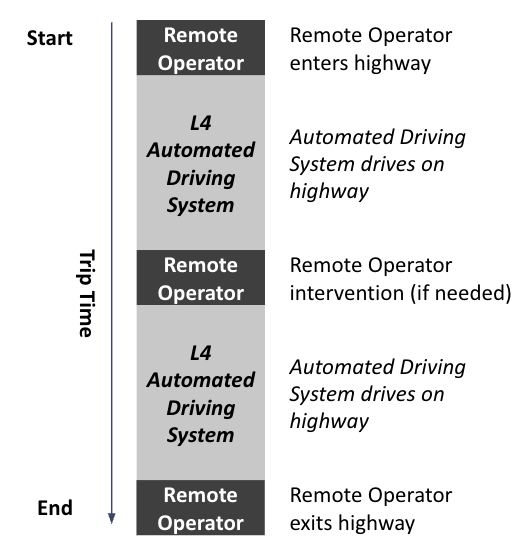}
    \caption{Simplified visualization of a Level 4 ADS hub-to-hub delivery use case with primarily driving on highways and limited exposure to surface streets between hub and highway}
    \label{Fig:.Remote_Operation_concepts}
\end{figure}

\subsection{Layer 1 - Road conditions}
\label{subsec:layer1}
The hub-to-hub use case includes mainly highways and limited surface streets. Highways are mostly stable in its characteristics, which simplifies the task for the ADS. Complexity increases at the transition between highway and surface street when using the highway on and off-ramps. ADS often handles the highway driving on its own while the first and last mile require support, as the surface street portion can vary in its characteristics and complexity.

RDS can perform this task with a controlled handover maneuver and transfer of the DDT between ADS and RDS. RAS, by its definition, cannot take over the DDT entirely, but can support ADS in a particular situation, which requires that an ADS in general can perform the DDT for a given ODD. To ensure a successful hub-to-hub mission with ADS and RAS, the ADS must generally be able to operate on all road types and the respective road infrastructure. Most ADS require adequate lane markings, which are given on highways except for temporary exclusions such as constructions, which will be discussed in layer 3 in \mbox{Section \ref{subsec:layer3}}. On surface streets, lane markers will also be mostly present, but there is a higher chance of gaps or misleading markers, which increases the need for remote support. An RDS can immediately take over the DDT and cover any distance without lane markers through human decision-making. In contrast, RAS can only provide limited input to the ADS by resolving a static situation, such as disengagements of the ADS due to a local problem. In addition, an RAS is not able to completely take over the DDT and close a longer gap in the lane markings, which makes continuous operation over several miles difficult.

\subsection{Layer 2 - Traffic infrastructure}
\label{subsec:layer2}
The traffic infrastructure poses a significant hurdle for ADS as the system has to perceive and correctly process traffic signs and traffic lights, which rarely occur on highways. In contrast, the speed limit on highways does not pose a general problem, as the travel speed of a truck generally remains within typical speed limits or even below. Similar to layer 1, only obstructions and temporary changes must be handled and are addressed in layer 3 in \mbox{Section \ref{subsec:layer3}}. On surface streets, these changes vary as the speed limits vary and therefore a reliable detection is essential. Further, traffic lights occur frequently and is therefore a core element for an ADS to handle. ADS can request RDS support in the following two forms: 

\begin{itemize}
    \item ADS performs an MRM in a situation which exceeds the safe ODD boundaries and let the RDS takeover from a \textbf{static situation}.
    \item ADS requests a \textbf{dynamic takeover} ahead of the critical situations.
\end{itemize}

The latter works well for a recurring, complex situation that ADS cannot handle alone, yet. In both scenarios, the RDS can resolve the situation through human decision-making and return the DDT to the ADS. In contrast, the RAS cannot take over the DDT entirely in any form. Thus all dynamic handover scenarios must be converted to static situations through an MRM. This may interrupt the mission progress and the flow of traffic. Once the ADS has created a static situation, the RAS can guide the ADS through the situation by providing additional input to the perception and decision making task. RAS can interpret traffic signs, traffic lights, and the overall traffic situation to suggest an appropriate dynamic driving strategy. A crucial distinction in the RAS design is whether it must contribute to the DDT by executing safety-critical tasks, particularly during time-critical phases of remote operation.

The second relevant traffic infrastructure element for this use case is the occurrence of weigh and toll stations, which represent highly predictable scenarios with clearly defined procedures that the ADS must be able to handle. The DDT itself does not pose significant additional complexity. The primary challenge lies in the interaction with onsite personnel. An ADS-only solution may require a predefined protocol to manage this interaction autonomously. However, if this protocol fails or unexpected deviations occur, a Remote Operator must intervene to ensure the continuity of the mission. In such cases, both RDS and RAS are capable of supporting the ADS, as the main task involves interpreting and responding to the personnel's on-site instructions. Given the static and well-structured nature of these interactions, RAS is generally sufficient, as it does not require full control capabilities such as pedals or steering inputs.

\subsection{Layer 3 - Temporal manipulations of layer 1 and 2}
\label{subsec:layer3}
As previously discussed by Hans and Walter \cite{hans2024backedautonomy}, the major challenge for ADS occurs on layer 3 as the ADS is generally well suited for most recurring situations in layer 1 and layer 2. Once a temporal manipulation occurs, the situation might be different. It is often hard to predict how such manipulations unfold, and with that, the variety of scenarios increases exponentially. Over time, ADSs will learn how to handle more of these situations, which reduces the need for remote support, but especially early on, ADS is reliant on remote support. Typical situations at layer 3 are construction sites and traffic accidents, where lane markers, traffic signs, and the general flow of traffic change. In constructions, the lane markings and traffic signs might be replaced with temporal markers, human sign holders, and signs that overrule the conventional signs. In case of traffic accidents, while the markers and signs are still present, the situation might require temporarily ignoring them.

The RDS can overtake the entire DDT and resolve this situation fairly well and fast because human intuition can understand and resolve the situation. Similar to prior layers, RAS can provide valuable input as well. However, its disadvantage is, that it can only do so from a static situation. To create such a static situation, the ADS has to first perform an MRM, which might lead to more complexity and impact for the traffic flow. RAS can guide the ADS through the situation via way points or basic perception decisions about traffic signs, lane markers, and maneuvers. While RDS support improves the system capabilities temporarily, the RAS support at best and brings the ADS back to its base performance. Therefore, RAS is more of a last resort to somehow get through the situation and continue the mission than an equivalent compensation for the ADS. 

\subsection{Layer 4 - Movable and dynamic objects}
\label{subsec:layer4}
At layer 4, the critical objects that contribute to determining challenges and support solutions are dynamic objects. ADS, as mentioned in layer 1 and layer 2, will handle most of the situations well. In situations where it reaches its performance limits, the ADS will typically trigger an MRM. The reason for that is that there is usually insufficient time for a handover to RDS immediately. The exceptions here are strategy decisions like lane selection. In that situation, the ADS may try to change lane to the priority lane, and if not successful, the RDS could take over and perform that lane change without mission interruption. RAS cannot meaningfully support the ADS in dynamic situations as it requires rather stationary conditions for input and support. In conclusion, the ADS must be able to resolve most layer 4 challenges independently of any remote support; otherwise, the system cannot perform its use case.

\subsection{Layer 5 - Environment conditions}
\label{subsec:layer5}
At layer 5, the primary limitation for ADS lies in low-visibility weather conditions such as darkness, heavy rain, snow, or fog. While radar-based or sensor-fusion systems offer greater robustness than camera-only setups, environmental conditions can significantly impair perception and reduce system availability. The support capabilities of RDS and RAS in these scenarios are similarly constrained, as both often rely on the ADS sensor stack or on camera-based inputs. However, RDS can partially compensate for reduced visibility, as human drivers are better at interpreting ambiguous situations such as occluded lane markings or snow-covered traffic lights. In contrast, RAS may assist only after the ADS has already disengaged and reached a static state. Its support is therefore limited to isolated interventions rather than continuous control in dynamic low-visibility scenarios. A second important use case at layer 5 is strategic route planning in response to predicted weather events. Here, RAS offers distinct advantages by assisting in rerouting decisions or advising temporary ODD adjustments. RDS, in contrast, is typically not configured for such planning tasks. Overall, layer 5 presents a mixed outcome: RDS is better suited for immediate, DDT-relevant interventions under adverse conditions, while RAS provides value in strategic planning and route adaptation.

\subsection{Layer 6 - Communication, digital info, network coverage}
\label{subsec:layer6}
The infrastructure layer plays a critical role in the selection between RDS and RAS. Unlike ADS, which can typically operate autonomously without continuous connectivity, both RDS and RAS depend on network availability although to different extents. RDS is designed for dynamic, time-critical interventions that require real-time control. As a result, it demands stable network conditions with high bandwidth and low latency to ensure safe and responsive operation \cite{hans2023operational}. Conversely, RAS typically supports the ADS in static situations, where the vehicle has already transitioned into an MRC. This allows for delayed or less time-sensitive interaction, resulting in significantly lower requirements for connectivity and network performance. In environments where network coverage is inconsistent, such as remote highway segments, the RAS is more resilient. Temporary drops in connection or latency spikes do not hinder its core functionality in the same way they would compromise the effectiveness of RDS. Therefore, in infrastructure-constrained areas, RAS offers a clear operational advantage.

\section{Conclusion and future work}
\label{sec:verANDval}
This section contextualizes the evaluation results with respect to the hypotheses outlined in the methodology. In addition, it outlines directions for future research.

\subsection{Conclusion}
\label{subsec:Results}
The evaluation confirms the hypotheses outlined in the methodology that neither RDS nor RAS is universally superior across all ODD layers. RDS demonstrates clear advantages in layer 3, where dynamic, DDT-intensive situations such as construction zones or accidents demand full control. This is expected, as RDS takes over the complete DDT, while RAS merely supports the ADS in a supervisory or advisory capacity. The required level of support also depends on the duration and complexity of the interaction, as longer or more involved engagements may exceed the capabilities of purely advisory systems. The required level of support also depends on the duration and complexity of the interaction, as longer or more involved engagements may exceed the capabilities of purely advisory systems. Importantly, the conscious choice between RAS and RDS as the remote support architecture can directly influence the overall system cost and complexity of the ADS stack, particularly in use cases like trucking, where full remote driving capabilities may not be necessary. In layer 5, results are mixed as the RDS is more effective in scenarios requiring active driving, such as low-visibility conditions, whereas RAS performs better in strategic tasks like route planning in response to weather constraints. Again, the distinction centers on the degree of DDT involvement. \mbox{Layer 6} favors RAS due to its lower technical demands regarding latency, bandwidth, and connectivity. RDS, by contrast, depends on stable, high-quality network infrastructure, which may not always be available, particularly in rural areas.

Ultimately, the evaluation underscores that the choice between RDS and RAS must be tightly aligned with the specific use cases and the performance boundaries of the ADS across the ODD layers. Systems with DDT limitations benefit more from RDS, while those requiring intermittent, strategic input are better served by RAS. Therefore, the structured use case analysis presented here offers a framework for informed decision-making and system selection.

\subsection{Future Work}
\label{subsec:future}
While this paper presents a structured framework for evaluating the suitability of RDS and RAS based on ODD-specific use cases, several important aspects remain open for further investigation. The current analysis is limited to a single use case for hub-to-hub highway transport by trucks. Future research should extend this evaluation to additional scenarios such as urban operation, last-mile delivery, or mixed-traffic environments to assess the generalizability of the findings across varied ODDs.

In addition, an important direction for future research is to explore the feasibility and safety implications of dynamic takeover by Remote Drivers during ongoing automated driving. Specifically, it remains to be investigated under which conditions a seamless and safe transfer of control is possible without the need to bring the vehicle to a full stop. Furthermore, it would be valuable to examine how Remote Drivers can learn from or adapt to such dynamic takeovers, possibly through repeated exposure, feedback mechanisms, or integrated performance monitoring, in line with initial findings from \mbox{Hans et al. \cite{hans2025identification, hans2025learning, hans2025evaluationremotedriverperformance}}.

Moreover, legal and regulatory aspects were intentionally excluded to maintain a purely use-case-driven and system-neutral analysis. However, real-world deployment is inherently constrained by legal requirements regarding remote intervention, operator responsibility, and system supervision. A systematic integration of these constraints is essential to align technical feasibility with legal applicability in future work.

\section*{Acknowledgment}
Ole Hans and Benedikt Walter, as the first authors, collectively contributed to this work. The authors want to thank Bogdan Djukic and Athanassios Lagospiris from Vay Technology for their invaluable support, insightful input, and constructive comments on this paper. Their contributions were instrumental in shaping the direction and findings of this research.

\bibliography{Reference}

\begin{thebibliography}{10}
\providecommand{\url}[1]{#1}
\csname url@samestyle\endcsname
\providecommand{\newblock}{\relax}
\providecommand{\bibinfo}[2]{#2}
\providecommand{\BIBentrySTDinterwordspacing}{\spaceskip=0pt\relax}
\providecommand{\BIBentryALTinterwordstretchfactor}{4}
\providecommand{\BIBentryALTinterwordspacing}{\spaceskip=\fontdimen2\font plus
\BIBentryALTinterwordstretchfactor\fontdimen3\font minus \fontdimen4\font\relax}
\providecommand{\BIBforeignlanguage}[2]{{%
\expandafter\ifx\csname l@#1\endcsname\relax
\typeout{** WARNING: IEEEtran.bst: No hyphenation pattern has been}%
\typeout{** loaded for the language `#1'. Using the pattern for}%
\typeout{** the default language instead.}%
\else
\language=\csname l@#1\endcsname
\fi
#2}}
\providecommand{\BIBdecl}{\relax}
\BIBdecl

\bibitem{schoitsch2016autonomous}
E.~Schoitsch, ``Autonomous vehicles and automated driving status, perspectives and societal impact,'' \emph{Information technology, society and economy strategic cross-influences (IDIMT-2016). 24th Interdisciplinary Information Management Talks}, vol.~45, no.~1, pp. 405--424, 2016.

\bibitem{mutzenich2021updating}
C.~Mutzenich, S.~Durant, S.~Helman, and P.~Dalton, ``Updating our understanding of situation awareness in relation to remote operators of autonomous vehicles,'' \emph{Cognitive research: principles and implications}, vol.~6, no.~1, pp. 1--17, 2021.

\bibitem{vay_teledriving}
O.~Hans, B.~Walter, and H.-L. Ross, ``Our teledriving-first approach: How we build teledrive technology around safety and the human driver,'' Vay Technology GmbH, Tech. Rep., 2023.

\bibitem{roboauto_teleoperation_2025}
{Roboauto}, ``Teleoperation – drive anything from anywhere,'' \url{https://roboauto.tech/solutions/teleoperation/}, 2025.

\bibitem{wachenfeld2016safety}
D.-I.~W. Wachenfeld and P.~Junietz, ``Safety assurance for highly automated driving--the {PEGASUS} approach,'' in \emph{Automated Vehicle Symposium San Francisco}, 2016.

\bibitem{diermeyer2011mensch}
F.~Diermeyer, S.~Gnatzig, F.~Chucholowski, T.~Tang, and M.~Lienkamp, ``{Der Mensch als Sensor-Der Weg zum teleoperierten Fahren},'' in \emph{AAET-Automatisierungssysteme, Assistenzsysteme und eingebettete Systeme f{\"u}r Transportmittel}, 2011, pp. 119--135.

\bibitem{fong2001vehicle}
T.~Fong and C.~Thorpe, ``Vehicle teleoperation interfaces,'' \emph{Autonomous robots}, vol.~11, pp. 9--18, 2001.

\bibitem{majstorovic2022survey}
D.~Majstorovi{\'c}, S.~Hoffmann, F.~Pfab, A.~Schimpe, M.-M. Wolf, and F.~Diermeyer, ``Survey on teleoperation concepts for automated vehicles,'' in \emph{2022 IEEE international conference on systems, man, and cybernetics (SMC)}.\hskip 1em plus 0.5em minus 0.4em\relax IEEE, 2022, pp. 1290--1296.

\bibitem{UNECE}
H.~F. in~International Regulations~for Automated Driving Systems (HF-IRADS), ``Human factors challenges of remote support and control: A position paper from {HF-IRADS},'' vol.~8, pp. 1--9, 2020.

\bibitem{BSIFlex1886}
{British Standards Institution}, ``{BSI Flex 1886: System Aspects for Remote Operation of Vehicles},'' London, UK, August 2023.

\bibitem{amador2022survey}
O.~Amador, M.~Aramrattana, and A.~Vinel, ``A survey on remote operation of road vehicles,'' \emph{IEEE Access}, vol.~10, pp. 130\,135--130\,154, 2022.

\bibitem{kettwich2021teleoperation}
C.~Kettwich, A.~Schrank, and M.~Oehl, ``Teleoperation of highly automated vehicles in public transport: User-centered design of a human-machine interface for remote-operation and its expert usability evaluation,'' \emph{Multimodal Technologies and Interaction}, vol.~5, no.~5, p.~26, 2021.

\bibitem{lu2022teleoperation}
S.~Lu, R.~Zhong, and W.~Shi, ``Teleoperation technologies for enhancing connected and autonomous vehicles,'' in \emph{2022 IEEE 19th International Conference on Mobile Ad Hoc and Smart Systems (MASS)}.\hskip 1em plus 0.5em minus 0.4em\relax IEEE, 2022, pp. 435--443.

\bibitem{brecht2024evaluation}
D.~Brecht, N.~Gehrke, T.~Kerbl, N.~Krauss, D.~Majstorovi{\'c}, F.~Pfab, M.-M. Wolf, and F.~Diermeyer, ``Evaluation of teleoperation concepts to solve automated vehicle disengagements,'' \emph{IEEE Open Journal of Intelligent Transportation Systems}, 2024.

\bibitem{AFGBV2022}
``{Verordnung zur Genehmigung und zum Betrieb von Kraftfahrzeugen mit autonomer Fahrfunktion in festgelegten Betriebsbereichen (Autonome-Fahrzeuge-Genehmigungs-und-Betriebs-Verordnung – AFGBV)},'' 07 2022, § 14 Anforderungen an die Technische Aufsicht.

\bibitem{bogdoll2022taxonomy}
D.~Bogdoll, S.~Orf, L.~T{\"o}ttel, and J.~M. Z{\"o}llner, ``Taxonomy and survey on remote human input systems for driving automation systems,'' in \emph{Advances in Information and Communication: Proceedings of the 2022 Future of Information and Communication Conference (FICC), Volume 2}.\hskip 1em plus 0.5em minus 0.4em\relax Springer, 2022, pp. 94--108.

\bibitem{SAEJ3016:2021}
{Society of Automotive Engineers}, \emph{{SAE J3016: Taxonomy and Definitions for Terms Related to Driving Automation Systems for On-Road Motor Vehicles}}, Std., April 2021.

\bibitem{neumeier2019teleoperation}
S.~Neumeier, P.~Wintersberger, A.-K. Frison, A.~Becher, C.~Facchi, and A.~Riener, ``Teleoperation: The holy grail to solve problems of automated driving? sure, but latency matters,'' in \emph{Proceedings of the 11th International Conference on Automotive User Interfaces and Interactive Vehicular Applications}, 2019, pp. 186--197.

\bibitem{schitz2021interactive}
D.~Schitz, G.~Graf, D.~Rieth, and H.~Aschemann, ``Interactive corridor-based path planning for teleoperated driving,'' in \emph{2021 7th International Conference on Mechatronics and Robotics Engineering (ICMRE)}.\hskip 1em plus 0.5em minus 0.4em\relax IEEE, 2021, pp. 174--179.

\bibitem{schimpe2022open}
A.~Schimpe, J.~Feiler, S.~Hoffmann, D.~Majstorovi{\'c}, and F.~Diermeyer, ``Open source software for teleoperated driving,'' in \emph{2022 International Conference on Connected Vehicle and Expo (ICCVE)}.\hskip 1em plus 0.5em minus 0.4em\relax IEEE, 2022, pp. 1--6.

\bibitem{brecht2024risk}
D.~Brecht and F.~Diermeyer, ``Risk-aware shared control for teleoperation of automated vehicles in dynamic environments,'' in \emph{2024 IEEE International Conference on Systems, Man, and Cybernetics (SMC)}.\hskip 1em plus 0.5em minus 0.4em\relax IEEE, 2024, pp. 3269--3274.

\bibitem{rohne2022implementing}
D.~Rohne, A.~Richter, and E.~Schwalb, ``Implementing {ODD} as single point of knowledge to support the development of automated driving,'' in \emph{2022 IEEE Int. Conf. on Systems, Man, and Cybernetics (SMC)}.\hskip 1em plus 0.5em minus 0.4em\relax IEEE, 2022, pp. 1364--1370.

\bibitem{sun2021acclimatizing}
C.~Sun, Z.~Deng, W.~Chu, S.~Li, and D.~Cao, ``Acclimatizing the operational design domain for autonomous driving systems,'' \emph{IEEE Intelligent Transportation Systems Magazine}, vol.~14, no.~2, pp. 10--24, 2021.

\bibitem{czarnecki2018towards}
K.~Czarnecki and R.~Salay, ``Towards a framework to manage perceptual uncertainty for safe automated driving,'' in \emph{Computer Safety, Reliability, and Security: SAFECOMP 2018 Workshops, ASSURE, DECSoS, SASSUR, STRIVE, and WAISE}, 2018, pp. 439--445.

\bibitem{koopman2019many}
P.~Koopman and F.~Fratrik, ``How many operational design domains, objects, and events?'' \emph{Safeai@ aaai}, vol.~4, 2019.

\bibitem{PAS}
{The British Standards Institution}, ``\BIBforeignlanguage{en}{{Operational Design Domain ({ODD}) Taxonomy for an Automated Driving System ({ADS})–Specification}},'' Standard BSI PAS 1883, 2020.

\bibitem{FDIS}
``\BIBforeignlanguage{en}{{Road Vehicles — Test scenarios for automated driving systems — Specification for operational design domain}},'' International Organization for Standardization, Standard ISO/FDIS 34503, 2020.

\bibitem{AVSC}
{Automated Vehicle Safety Consortium}, ``{AVSC} best practice for describing an operational design domain: Conceptual framework and lexicon,'' \emph{SAE Industry Technologies Consortia}, 2020.

\bibitem{hans2023operational}
O.~Hans, M.~Avezum, S.~Borysov, H.-L. Ross, and J.~Adamy, ``Operational design domain qualification framework for remotely driven vehicles in urban environment,'' in \emph{2023 IEEE International Automated Vehicle Validation Conference (IAVVC)}.\hskip 1em plus 0.5em minus 0.4em\relax IEEE, 2023, pp. 1--6.

\bibitem{hans2024backedautonomy}
O.~Hans and B.~Walter, ``{ODD} design for automated and remote driving systems: A path to remotely backed autonomy,'' in \emph{2024 IEEE the 9th International Conference on Intelligent Transportation Engineering (ICITE)}.\hskip 1em plus 0.5em minus 0.4em\relax IEEE, 2024.

\bibitem{hans2025identification}
O.~Hans and J.~Adamy, ``Identification and classification of human performance related challenges during remote driving,'' \emph{arXiv preprint arXiv:2503.09865}, 2025.

\bibitem{hans2025learning}
{Hans, Ole and Adamy, Jürgen}, ``Learning from disengagements: An analysis of safety driver interventions during remote driving,'' in \emph{2025 IEEE Intelligent Vehicles Symposium (IV)}.\hskip 1em plus 0.5em minus 0.4em\relax IEEE, 2025.

\bibitem{hans2025evaluationremotedriverperformance}
O.~Hans, B.~Walter, and J.~Adamy, ``Evaluation of remote driver performance in urban environment operational design domains,'' \emph{IEEE Open Journal of Intelligent Transportation Systems}, vol.~6, pp. 722--737, 2025.

\end{thebibliography}

\end{document}